# Metallisation of the Mott Insulator Ca$_2$RuO$_4$ using Electric Double-Layer Gating

Tatsuhiro Sakami[1], Hiroki Ogura[1], Akihiro Ino[1], Takumi Ouchi[2], Tsutomu Nojima[2], and Fumihiko Nakamura[1]*

[1]*Kurume Institute of Technology, 2228-66 Kamitsu, Kurume 830-0052, Japan*
[2]*Institute for Materials Research, Tohoku University, Sendai 980-8577, Japan*

To verify whether the Mott insulator Ca$_2$RuO$_4$ can be switched by applying electric-field alone, regardless of current flow, we employ metallisation using electric double-layer gating (EDLG). The resistance change due to EDLG occurs only when positive gate-voltage above +3 V is applied. The amplitude of the reduction, reaching ~97% of the initial value, is difficult to interpret as surface metallisation and is likely related to structural change in bulk.

The Mott insulator Ca$_2$RuO$_4$ (CRO) can be switched to metal by applying relatively weak electric-field ($E$-field) of ~40 V/cm at 295 K.[1] This switching (SW) is one of the most attractive phenomena not only in condensed matter physics but also in device physics. We can expect that the SW in CRO allows us to develop the Mott device operating at room temperature. However, there remains argue of whether insulating CRO can be switched by applying $E$-field alone regardless of the influence of Joule heating (current flowing).[2,3] This is because Joule heating sometimes prevents us to know the intrinsic nonlinear conduction in the SW phenomana.[2,3] In order to drive away the influence of Joule heating from the SW in CRO, an electrostatic method, represented by "electric double-layer gating (EDLG)" is suitable. This technique is known as a powerful tool for metallising various types of insulators including usual band insulators and the Mott insulator even in the absence of Joule heating.[4-9] In this study, we adopted the EDLG technique to examine the pure $E$-field effect on CRO.

Figure 1(a) illustrates an Electric Double Layer Transistor (EDLT) device fabricated using single-crystalline CRO. A gate bias $V_G$ from −4 to +4 V was applied to the freshly cleaved (001) surface of CRO crystals using a Pt gate electrode via an ionic liquid (N, N-diethyl-N-(2-methoxyethyl)-N-methylammonium bis(trifluoromethylsulphonyl)imide: DEME-TSFI).[10] The single crystals of CRO were grown by the floating-zone method with RuO$_2$ self-flux using a commercial infrared furnace (Canon Machinery, model SC-M15HD).[1] Mainly, we studied the EDLT using the crystals with exactly stoichiometric oxygen content, which were stably grown under the optimal atmospheric conditions (a gas mixture of 10% oxygen and 90% argon at total pressure of 1.0 MPa). It is well known that the growth of single-crystals with the non-stoichiometric oxygen content is considerably difficult. After much effort, we successfully obtained a few crystals of $Ca_2RuO_{4+\delta}$ with excess oxygen content ($\delta$ ~0.02) under the growth conditions of high partial oxygen pressure of 20-60% in an oxygen-argon gas mixture at 1.0 MPa,[11-13] where the excess oxygen was considered to locate only at the surface, as discussed later. We also prepared EDLT using $Ca_2RuO_{4+\delta}$ as a reference sample. In contrast, the oxygen-deficient CRO crystals have never been obtained because the crystal growth becomes entirely unstable under conditions of the low partial oxygen pressure less than 0.1 MPa.

Figure 1(b) shows the time series of resistance ($R$) with varying $V_G$ from −4.0 to +4.0 V for sample #1. The measurement was performed for stoichiometric CRO at a constant temperature of 260 K under helium gas atmosphere (in a helium vessel). In order to clarify the $V_G$ variation of the resistance change, the data on the ratio of resistance change were replotted in Fig. 1(c) as a function of $V_G$ at 260 K, indicated by red dots. While the resistance remains constant in the $V_G$ range from −4.0 to +3.0 V (before 2500 s), a rapid reduction in $R(V_G)$ is induced by applying $V_G \geq$ +3.0 V (after 2500 s). Moreover, the resistance decreases initially at a rate of $\frac{1}{R_0}\left(\frac{\Delta R}{\Delta t}\right) = -3 \times 10^{-4}$ s$^{-1}$ (where $R_0$ is 900 Ω for sample #1) when $V_G$ is increased to the upper limit of +4.0 V. The reduction is progressive over time whereas the rate gradually decreases. It is noteworthy that the resistance reduction can be induced only by applying a "positive" $V_G$ above +3.0 V. This indicates that metallisation of insulating CRO cannot be induced by hole doping but requires electron doping. We emphasise here that the resistance reduction induced by using EDLG provides direct evidence that metallisation is solely induced by applying $E$-field, even in the absence of Joule heating.

Figures 1(d) and (e) show that the extent and the reproducibility of resistance reduction for sample #2 (stoichiometric CRO). The measurement was performed at 260 K under helium gas atmosphere. The time series of resistance and $V_G$ are plotted in top and bottom views, respectively. When $V_G$ was maintained at +4.0 V for 24 h after increasing $V_G$ up to +4.0 V (Fig. 1(d)), the resistance continued to reduce reaching ~35% of the initial value (5 kΩ for sample #2). Subsequently, when we return $V_G$ from +4.0 to 0 V and maintain it in this state for further 24 h (Fig. 1(e)), the resistance gradually recovered to ~95% of its initial value. We can, thus, confirm good reversibility in the resistance change induced by applying $V_G$.

It is difficult to interpret the reversible resistance change as an electrochemical reaction, such as hydrogen (proton) migration or oxygen reduction. As for the hydrogen migration, the resistance reduction was observed only when dehydration treatment of the ionic liquid was conducted thoroughly using an evacuated oven at 120°C. In contrast, using an ionic liquid suspected to have absorbed moisture prevents us from observing the resistance reduction, indicating the importance of removing residual moisture within the ionic liquid to induce the phenomenon. We, therefore, exclude the possibility of hydrogen (proton) migration. As for the oxygen reduction, it is quite difficult to create oxygen defects in CRO as mentioned before. Even if it happens, it is unlikely that the oxygen ions diffused from the crystal into the ionic liquid by applying positive $V_G$ would fully return to the crystal simply by resetting $V_G$ to 0 V without applying negative $V_G$, as reported for $YBa_2Cu_3O_y$-EDLT.[9)] In addition, the EDLT with excessively oxygenated CRO does not show the resistance change, as shown in Fig. 1(c) (sample #3: black open circles). According to the recent angle-resolved photoemission spectroscopy, the impact of introducing excess oxygen is limited to the electronic states at the surface of the CRO crystals.[11)] Our EDLG experiments for sample #3 results also support that the oxygen cannot move by applying $V_G$, because the excess oxygens are known to affect the conductance at the interface. With the reasons mentioned above, it can be concluded that the effect of resistance reduction is a purely electrostatic phenomenon, distinct from an electrochemical reaction.

We note that the resistance reduction of stoichiometric CRO crystals is induced not only by increasing $V_G$ but also by the aging effect; i.e., the resistance continues to decrease over time under a constant $V_G$. To examine this, EDLG experiments were performed in a vacuum chamber equipped with a GM refrigerator over a long period (sample #4). Figure 2(a) shows the resistance changes plotted as a function of time (days) over a period of nine months. Here, we continued to measure the resistance when $V_G$ was increased up to +4 V over 5 days at 260 K and then maintained at +4 V for a long duration from 5 to 285 days. Similar to the result in Fig. 1(b), the resistance begins to reduce when $V_G$ is applied above +3 V, indicating that the vacuum and helium gas environments are equivalent for the operation of our EDLT device. The resistance, initially 700 Ω ($V_G < +3$ V), decreases to ~ 600 Ω on the 5th day after applying $V_G$ = +4.0 V. Moreover, it continues to decrease over time, reaching 18 Ω, corresponding to an ~97% reduction, after 285 days. After this, the resistance started to increase gradually by changing $V_G$ to zero. A few months later, the resistance of the CRO crystal in the EDLT device, which was retrieved from the vacuum chamber, was ~750 Ω (at ~300 K), indicating that it had almost recovered to its initial value of ~670 Ω. Thus, we confirm the reversibility of the resistance reduction due to the aging effect.

Next, let us evaluate the metallic nature of the resistance from its $T$ variation. Figure 2(b) shows $R(T)$ curves measured at several fixed $V_G$. The $R(T)$ curves at $V_G \leq +3.0$ V is roughly interpreted as the activation type conduction with the energy gap of 0.2 eV, which is consistent with previous report.[14)] With increasing $V_G$ from +3.0 to +4.0 V the slope of $R(T)$ curve (*i.e.*, $T$ coefficient, $dR/dT$) becomes gradual. The most interesting finding here is that the negative slope becomes more gradual with aging variation when we have maintained at $V_G = +4.0$ V for long periods from 6 to 285 days. That is to say that the coefficient $dR/dT$, which is initially $-30$ Ω/K immediately after applying $V_G = +4.0$ V, decreases to $-0.35$ Ω/K after 180 days. Although our current measurements show every $R(T)$ curve remains with a negative coefficient ($dR/dT < 0$), improvements in the EDLT device, such as sample size and mounting methods, will allow us to metallise the insulating CRO in the near future.

Here, we will focus on evaluating the progress of the EDLG-induced metallisation in CRO; however, there is currently a great lack of reliable information on the metallisation process in

CRO. This is because the metallisation in CRO is strongly related to its complex and unique first-order phase transition, which is accompanied by a large and anisotropic structural change, a negative thermal expansion and a mixed state where the metallic and insulating phases coexist. We, therefore, attempt to evaluate the depth of the metallic region, based on a simple model where the metallisation induced by EDLG progresses uniformly from the surface to the inside, comparing the initial resistance of ~700 Ω with the final value of ~18 Ω. We used here the sample with a thickness of 0.1 mm. Furthermore, as reported in our previous works, the resistivity at 260 K is 10 Ω·cm for the non-metallic state and ~250 mΩ·cm for the metallic state.[15, 16] From these parameters, we can estimate the thickness of metallic region in CRO is ~100 nm. This value is too thick to be interpreted as a surface metallisation effect observed in typical oxide insulators; the thickness of a metallic region formed on the surface of conventional band insulator using EDLG has been reported to be ~10 nm for $SrTiO_3$ and ~1 nm for $MoS_2$.[17, 18] These results indicate that the EDLG-induced metallisation in CRO is hard to explain in terms of a conventional electrostatic carrier doping or injection only at the sample surface.

We infer that strong electron-correlation plays an important role in the metallisation of CRO induced by EDLG. Indeed, it has been reported that the changes in the surface electronic state can spread into the bulk with ~100 nm in case of strongly correlated systems such as $VO_2$ and manganites.[7, 8] In the case of single-crystalline CRO, the EDLG induced phenomena are characterised by long relaxation times and slow progression, which to our knowledge are quite rare. Considering these with the fact that CRO is sensitive to structural changes in metallisation through the strong electron-lattice coupling,[1, 19, 20] the enormous resistance changes with the slow dynamics observed in this study are most likely due to the bulk-like structural change extending into the crystal interior, which is induced by gating. Especially, the lattice deformation can proceed more slowly when the crystals are fixed to the substrates as in our devices.[21]

We can fully expect that the initial surface metallisation induced by EDLG triggers structural changes and further promotes metallisation. In other words, a cascading of metallisation and structural change induces the bulk metallisation. Based on our inference, we can also account for the resistance reduction of CRO maintained for as long as 285 days under EDLG. To validate our inference, it is required to measure the surface strain of CRO crystals by using X-ray diffraction under EDLG.

We acknowledge Y. Koga and H. Nakajima for their experimental assistances. This work was supported by Grant-in-Aid for Scientific Research (Grant Nos. 17H06136, 20K03842, 22K03485 and 23K22437) by JSPS.

*E-mail: fumihiko@kurume-it.ac.jp

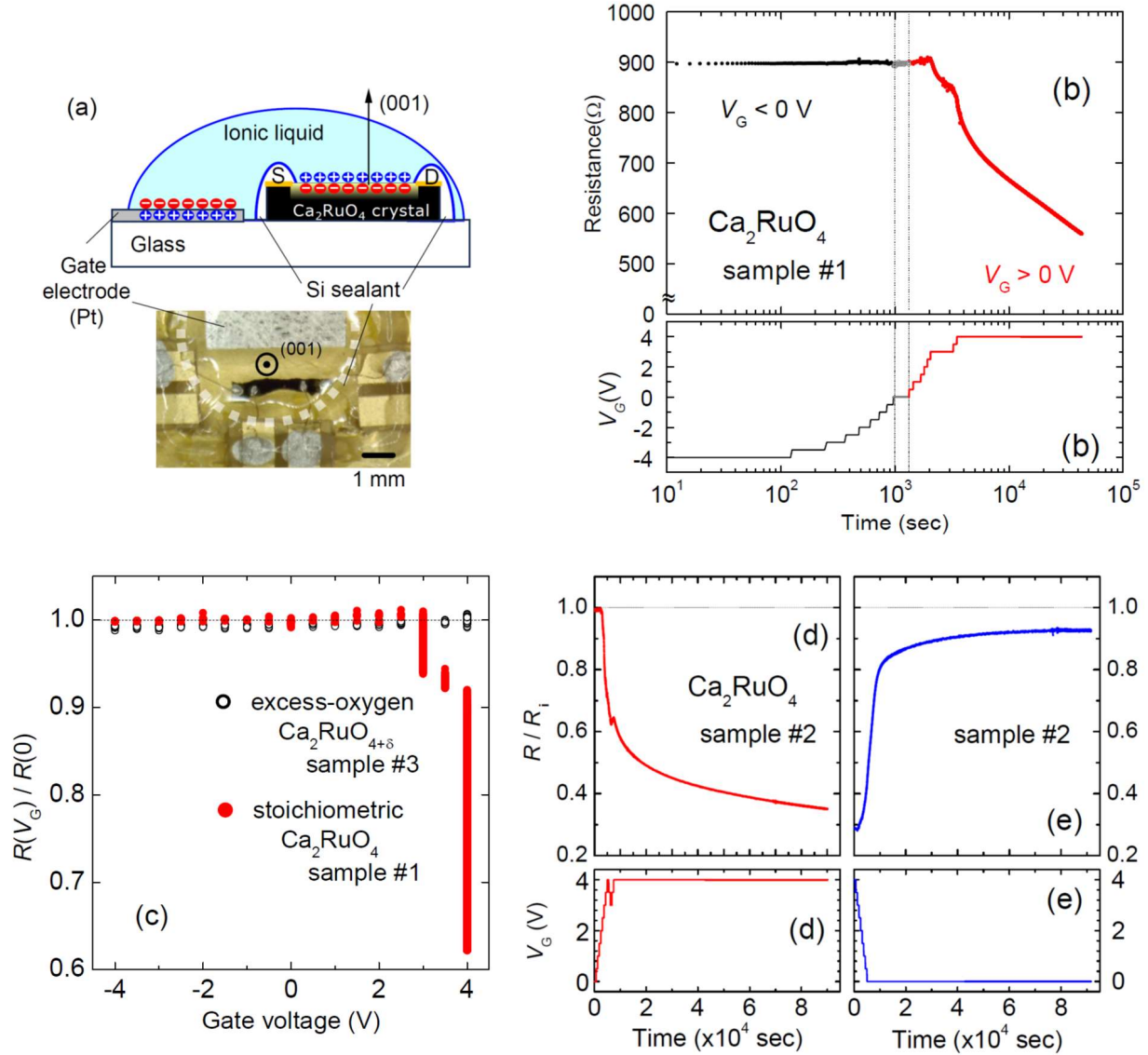

**Fig. 1.** (colour online) (a) A schematic cross-section drawing (left) and top-view photo (right) of an EDLT device fabricated using single-crystalline $Ca_2RuO_4$. $V_G$ was applied on the freshly cleaved (001) surface via an ionic liquid. The typical size of our crystal in this work was 2 ~ 3 mm long, 0.3 ~ 0.5 mm wide and 0.05~ 0.1 mm thick. (b) The time series (in seconds) of changes in resistance (red) and varied $V_G$ (blue broken line) for sample #1. The measurement was performed at 260 K, maintained in a helium vessel. (c) The change of resistance ratio $R(V_G)/R(0)$ at 260 K plotted as a function of $V_G$. The red closed circles show the data for stoichiometric CRO (sample #1) at 260 K, while the black open circles show the data for excessively oxygenated CRO at 260 K, maintained by using a GM refrigerator (sample #3). (d, e) The time series (in seconds) of the ratio of resistance, $R/R_i$ (top) and $V_G$ (bottom), for sample #2 where initial resistance at 260 K is 5 kΩ. The processes of increasing and decreasing $V_G$ are represented in (d) red and (e) blue, respectively.

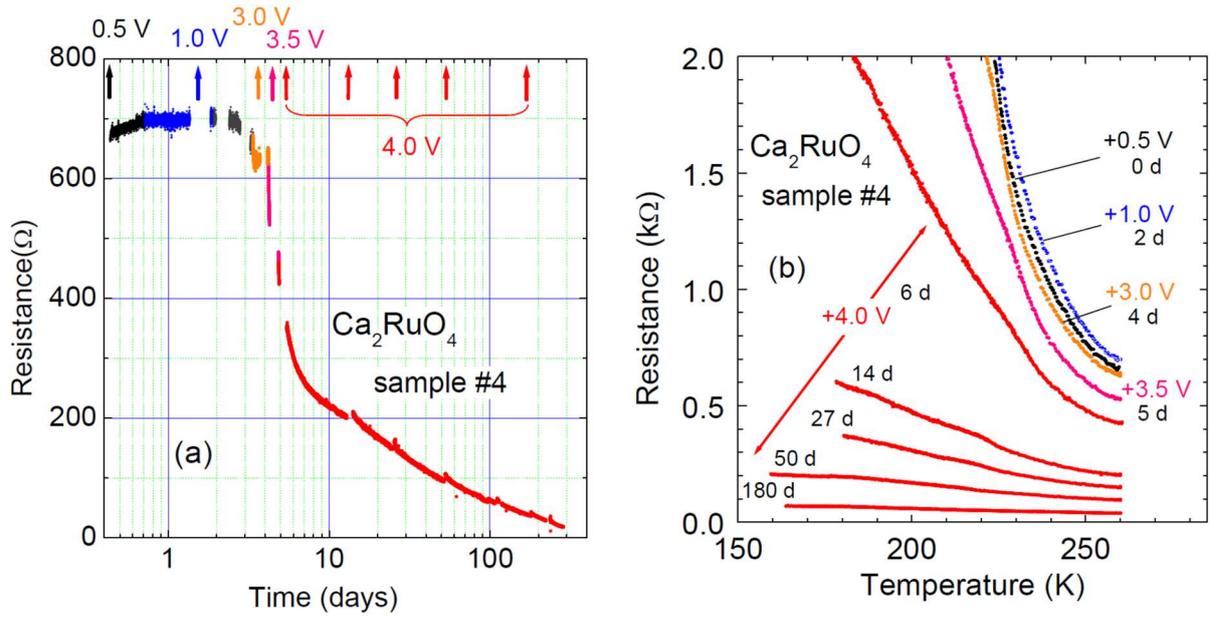

**Fig. 2.** (colour online) (a) Variation in the resistance at 260 K for sample #4 plotted as a function of time (up to 284 days). The arrows with gate voltages indicate the timing at which $R(T)$ curves were measured. The difference in $V_G$ is represented by colour cording. (b) $R(T)$ curves measured at several fixed $V_G$ of +0.5, +1.0, +3.0, +3.5 and +4.0 V. These curves are also represented by the same colour-coding as in (a) for each $V_G$. Temperature coefficients of the resistance (d$R$/d$T$) start to become more gradual above $V_G$= +3.5 V. When $V_G$ = +4.0 V is continuously applied for a long period of time, the coefficient d$R$/d$T$ decreases from an initial value of −30 to −0.35 Ω/K after 180 days.